\documentstyle[12pt]{article}
\newcommand{\oB}{\vert_{\partial M}}
\begin{document}
\begin{flushright}{IC/94/373}\\
{gr-qc/9411062}\end{flushright}
\centerline{\large \bf Heat kernel for
antisymmetric tensor field on a disk}
\centerline{Dmitri V.Vassilevich}

International Centre for Theoretical Physics, Trieste, Italy

\centerline{and}

Department of Theoretical Physics, St. Petersburg University,

198904 St.Petersburg, Russia \footnote {Permanent addres, e.-mail:
dvassil @ sph.spb.su}

\vspace{3mm}

\centerline{Abstract}

We suggest a method of reduction of mixed absolute and
relative boundary conditions to pure ones. The case of
rank two tensor is studied in detail. For four-dimensional
disk the corresponding heat kernel is expressed in terms of
scalar heat kernels. The result for scaling behavior $\zeta (0)$
agrees with previous calculations.

\vspace{3mm}

{\bf 1. Introduction}

The modern interest to heat kernel expansion for manifolds with
boundaries is connected with applications to quantum cosmology
and quantum state of the universe (see monograph [1] and references
therein). This topic is also studied by mathematicians, especially
when related to $p$-forms and de Rham complex. It is worth noting
that the absolute and relative boundary conditions, which are
natural for the de Rham complex [2], are also singled out by the
BRST-invariance of quantum theory [3,4]. The absolute boundary
conditions are defined [2] as the Neumann boundary conditions for
tangential components of a $p$-form and the Dirichlet conditions
for other components
$$\partial_0 B_{i...k}\oB =0,\quad B_{0i...k} \oB =0 \eqno (1)$$
The relative boundary conditions are dual to the set (1). These
conditions are mixed. There were contradictions in computations
of the heat kernel expansion for mixed boundary conditions even
for QED in covariant gauge [3,5]. These contradictions were
removed recently [6,7] using the corrections [7] to the
analytic expressions [8]. However, further checks are highly
desirable.

The aim of this work is two-fold. First, we suggest a procedure
of reduction of the mixed boundary conditions on a disk to pure
ones and derive expression for all the Seeley-Gilkey coefficients
in terms of the heat kernel expansion for scalar field. This
expression is especially efficient in analyzing higher terms in the
heat kernel expansion. Second, in a particular case
we compare our results with calculations [6] based on analytic
formulae [8,7] and find that these two approaches agree. To
achieve these goals we do the following. We analyze the Laplace
operator for $p$-forms on $d+1$-dimensional disk and find that
the spectrum can be expressed in terms of $p$, $p-1$,...,0-forms
with tangential indices only satisfying pure boundary conditions.
We find that the degeneracies $D_n$ and the eigenvalues
$-\lambda_n^2$ have the same form as for scalar fields and express
the heat kernel
$$K(t)=\sum_n D_n exp (-\lambda_n^2)$$
in terms of the scalar heat kernels. The case of two-forms
is analyzed in detail.

\vspace{3mm}

{\bf 2. Harmonic expansion}

Consider $d+1$ dimensional unit disk with the metric
$$ds^2=dr^2+r^2 d\Omega^2, \quad 0 \le r \le 1 \eqno (2)$$
where $d\Omega^2$ is the metric on unit sphere $S^d$.
Throughout this paper we use the notations $\{ x_\mu \} $$=
\{ x_0,x_i\}$, $x^0=r$, $\mu =0,1,...,d$. The $d+1$
dimensional Laplace operator $\Delta =\nabla^\mu
\nabla_\mu$ acting on $p$-form $B$ can be written as
$$(\Delta B)_{i_1...i_{p-1}0}= \left ( \partial_0^2+
\frac {d-2p+2}r \partial_0 +\frac {p^2-dp-1}{r^2}
+\ ^{(d)}\Delta \right ) B_{i_1...i_{p-1}0}-$$
$$-\frac 2r \ ^{(d)}\nabla^k B_{i_1...i_{p-1}k} \eqno (3)$$
$$(\Delta B)_{i_1...i_p}= \left ( \partial_0^2+
\frac {d-2p}r \partial_0 +\frac {p^2-dp}{r^2}
+\ ^{(d)}\Delta \right ) B_{i_1...i_p}+$$
$$+\frac 2r \sum_{a=1}^p \ ^{(d)}\nabla_{i_a}
B_{i_1...i_{a-1}0i_{a+1}...i_p} , \eqno (4)$$
where $\ ^{(d)}\nabla$ and $\ ^{(d)}\Delta$ are the
covariant derivative and Laplace operator corresponding
to $d$-dimensional metric $g_{ik}$.

We assume that the $p$-from $B$ satisfy absolute or
relative boundary conditions. Let us outline the strategy
of finding the spectrum of the Laplace operator (3), (4).
First we make the Hodge-de Rham decomposition, which is
orthogonal for the boundary conditions in question.
Since the metric (2) is flat, the spectrum of the Laplace
operator on longitudinal $p$-forms is identical to that
on the transversal $p-1$-forms. This is a well known fact
from the cohomology theory. Hence at the first step our
problem is reduced to two eigenvalue problems for
transversal $p$ and $p-1$ froms. We shall assume that all
the cohomology groups are trivial (this is indeed restriction
on $p$). The modifications for the case of non-trivial
cohomology are straightforward.

Consider the $d+1$ dimensional tarnsversality condition
$$\nabla^\mu B_{\mu \nu ...\rho}=0 . \eqno (5)$$
On a disk it can be written as
$$(\nabla B)_{i_1...i_{p-2}0}=\ ^{(d)}\nabla^i
B_{ii_1...i_{p-2}0}=0 \eqno (6)$$
$$(\nabla B)_{i_1...i_{p-1}}=(\partial_0+
\frac {d-2p+2}r )B_{0i_1...i_{p-1}}+
\ ^{(d)}\nabla^i B_{ii_1...i_{p-1}} =0 \eqno (7)$$
Of course, for $p=1$ the condition (6) is absent. For
$p=d+1$ the theory is easily redused to the scalar one
with pure boundary conditions. For $p \le d$ the
equations (6), (7) have two types of solutions depending
on whether $B_{i_1...i_p}$ is transversal or longitudinal
as a $p$-form on $S^d$. For the solutions of the first
type $B_{0i_1...} $ is identicaly zero. For the solutions
of the second type both $B_{0i_1...}$ and $B_{i_1...i_p}$
can be expressed via transversal $p-1$-form on $S^d$.
The eigenvalue problem for the second type
solutions is the same
as for the first type but with $p-1$-form instead of
$p$-form. Thus as a result of the second step the
problem is reduced to the eigenvalue equation for
operator (4) acting on anisymmetric transversal tensors
on $S^d$. Note, that now these tensors satisfy pure
boundary conditions if either absolute or relative
boundary conditions were choosen initially. The corresponding
eigenfunctions can be expressed in terms of Bessel
functions. The eigenvalues are defined by zeros of these
functions or their derivatives.

Furthermore, the heat kernels can be related to that for
scalar fields. Eigenfunctions are tipically proportional
to $J_n(\lambda r)Y^n(x_j)$ for odd $d$ and to
$J_{n+\frac 12} (\lambda r)Y^n(x_j)$ for even $d$,
where $J_n$ are the Bessel functions and $Y^n$ are
tensor spherical harmonics. The eigenvalues $\lambda$
are defined by boundary conditions. The degeneracies
$D_n$ for odd $d$ are polynomials of $n^2$. For even $d$
the $D_n$ contain only odd powers of $n+\frac 12$. This
structure is common for all $p$-forms. This fact allows us
to express heat kernel for $p$-forms
in terms of scalar heat kernels in $d$, $d-2$, etc. dimensions.

To illustrate this procedure let us consider the case of
$p=2$. The absolute and relative boundary conditions are
respectively
$$\partial_0 B_{ik} \oB = (\nabla_0 +\frac 2r )B_{ik} \oB
=0 , \quad B_{oi} \oB =0 \eqno (8A)$$
$$(\partial_0 +\frac {d-2}r )B_{i0} \oB =
(\nabla_0 +\frac {d-1}r )B_{i0} \oB =0 , \quad
B_{ik} \oB =0 \eqno (8R)$$
One can see that for longitudinal forms $B_{\mu \nu}^L$$=
\nabla_\mu A_\nu -\nabla_\nu A_\mu$ the corresponding
1-forms $A$ also satisfy absolute or relative boundary
conditions
$$\partial_0 A_i \oB = (\nabla_0 +\frac 1r )A_i \oB =0,
\quad A_0 \oB =0 \eqno (9A)$$
$$(\partial_0 +\frac dr )A_0 \oB=(\nabla_0 +\frac dr) \oB =0,
\quad A_i \oB =0 \eqno (9R)$$
This follows from the general theory [2] and can be verified
by direct calculations. It is obvious that
$$\Delta (\nabla_\mu A_\nu -\nabla_\nu A_\mu )=
\nabla_\mu \Delta A_\nu - \nabla_\nu \Delta A_\mu \eqno (10)$$
Hence the spectrum of the Laplace operator $\Delta$ on
$B^L_{\mu \nu}$ is defined by the spectrum of the Laplace
operator on transversal vector fields $A^T_\mu$.

Consider now transversal 2-forms. As it was mentioned earlier,
the equations (6), (7) have solutions for which $B_{0i}=0$ and
$B_{ik}$ is arbitrary transversal 2-form on $S^d$,
$\ ^{(d)}\nabla^i B_{ik}=0$. Such modes decouple from other
modes and satisfy pure boundary conditions (see (8A) and (8R)).
The corresponding eigenvalue problem will be considered in the
next section.

After some algebra one can find the rest of the solutions of (6)
and (7).
$$B_{ik}(\phi )=(\partial_0 +\frac {d-4}r )r
(\partial_i \phi_k -\partial_k \phi_i ),$$
$$B_{i0}(\phi )=(\ ^{(d)}\Delta -\frac {d-1}{r^2} )r \phi_i,
\eqno (11)$$
where $\phi_i$ is arbitrary transversal vector on $S^d$,
$\ ^{(d)}\nabla^i\phi_i =0$. After lengthy but straightforward
calculations one obtains
$$\Delta B_{\mu \nu} (\phi )=B_{\mu \nu }(\Delta \phi )
\eqno (12)$$
The boundary conditions for $\phi$ leading to (8A) and (8R) are
respectively
$$\phi_i \oB =0 \eqno (13A)$$
$$(\partial_0 +\frac {d-3}r )\phi_i \oB =
(\nabla_0 +\frac {d-2}r )\phi_i \oB =0 \eqno (13R)$$
To prove that the boundary conditions (8A) and (8R) are
ensured by (13A) and (13R) it is useful to expand $\phi_i$
in a sum of harmonics which are eigenvalues of both
$\Delta$ and the Laplace operator on unit $S^d$, $\tilde \Delta
=r^2\ ^{(d)}\Delta$. Equations (13R) and the second equation
of (13A) become evident. Note a helpful operator identity
$$\partial_0\ ^{(d)}\Delta =\ ^{(d)}\Delta
(\partial_0 -\frac 2r ). \eqno (14)$$
For the first of the conditions (8A) we have
$$\partial_0 B_{ik}(\phi ) \oB =\left (\ ^{(d)}\nabla_i r
(\partial_0 +\frac 1r )(\partial_0 +\frac {d-3}r )\phi_k
-(i \leftrightarrow k) \right ) \oB =$$
$$=\left ( \ ^{(d)}\nabla_i r (\Delta -\frac {\tilde \Delta}{r^2}
+\frac {d-1}{r^2} ) \phi_k - (i \leftrightarrow k) \right ) \oB
\eqno (15)$$
where we used transversality of $\phi_k$ and the equation (4)
for $p=1$. If $\phi_k$ satisfy Dirichlet boundary condition
(13A), the last line vanishes term by term for every eigenmode
of $\Delta$ and $\tilde \Delta$.

It is important to note that the orthogonal harmonics of $\phi_i$
generate orthogonal harmonics of $B_{\mu \nu}(\phi )$. In
both cases we use standard inner product without surface
terms.

Now we are to reduce the harmonic expansion for $A^T_\mu$ to
pure boundary condition problem. This was done in Ref. [7],
thus here we give the results only. By solving the transversality
condition (7) we arrive again at two types of solutions. The one
is described by $A_0=0$, $\ ^{(d)}\nabla^i A_i=0$ with pure
boundary conditions (9A) and (9R) for $A_i$, and the other
is expressed in terms of scalar field $\psi$
$$A_0(\psi )= ^{(d)}\Delta r\psi , \quad A_i(\psi )=
-\ ^{(d)}\nabla_i (\partial_0 +\frac {d-2}r )r\psi .
\eqno (16)$$
The boundary conditions for $\psi$ are given by the
following equations
$$\psi \oB =0 \eqno (17A)$$
$$(\partial_0 +\frac {d-1}r )\psi \oB =0 \eqno (17R)$$
for absolute and relative boundary conditions on $A_\mu$
respectively. The decomposition commutes with the
Laplace operator, $\Delta A_\mu (\psi )= A_\mu (\Delta \psi )$,
and satisfy all necessary orthogonality properties.

In this section we reduced the harmonic expansion for the
2-form $B_{\mu \nu}$, satisfying absolute or relative mixed
boundary conditions, to the expansions of the fields
$B_{ik}^T$, $\phi_i^t$, $A_i^T$ and $\psi$. All of them
are subject to pure boundary conditions.

\vspace{5mm}

{\bf 3. Heat kernel expansion}

Let us express the heat kernel for 2-form in $d+1=4$
dimensions via scalar heat kernels. First recall the
eigenvalues $a_l$ and degeneracies $D_l$ of the Laplace
operator $\tilde \Delta$ on unit $S^3$ acting on scalars,
transversal vectors and transversal two-forms.
$$a_l^0=-l(l+2), \quad D_l^0=(l+1)^2, \quad l=0,1,...$$
$$a_l^1=-l(l+2)+1, \quad D_l^0=2l(l+2), \quad l=1,2,...$$
$$a_l^2=-l(l+2)+2, \quad D_l^2=(l+1)^2, \quad l=1,2,...
\eqno (18)$$
As it was mentioned before, the transversal 3-dimensional
harmonics decouple from other components of corresponding
tensors. By substituting (18) in (4) we obtain the
eigenfunctions of the Laplace operator for scalar, vector
and tensor fields respectively
$$ r^{-1}J_{l+1}(\lambda r)Y^l(x_j), \quad
J_{l+1}(\lambda r)Y^l_i(x_j), \quad
rJ_{l+1}(\lambda r)Y^l_{ik}(x_j), \eqno (19)$$
where $Y^l$, $Y^l_i$ and $Y^l_{ik}$ are spherical harmonics.
We used $\ ^{(3)}\Delta =\frac 1{r^2} \tilde \Delta$.
The eigenvalues $-\lambda^2$ are defined by boundary conditions.

Denote by $S$ the number appearing in the Neumann boundary
conditions for some field $\Phi$
$$(\nabla_0 -S) \Phi \oB =0 \eqno (20)$$

Consider absolute boundary conditions. The integrated heat kernel
for the field $B^T_{ik}$ has the form
$$K_B(t)=\sum_{l=1}^\infty \sum_{\lambda_l} (l+1)^2
\exp (-\lambda_l^2 t)=$$
$$=\sum_{l=0}^\infty \sum_{\lambda_l} (l+1)^2
\exp (-\lambda_l^2 t) -\sum_{\lambda_0} \exp (-\lambda_0^2 t)
\eqno (21)$$
The eigenvalues $\lambda$ are defined by the condition
$\partial_0 r J_{l+1} (r\lambda )\vert_{r=1}=0$. This condition
is equivalent to $(\partial_0 +2) r^{-1} J_{l+1}
(r\lambda )\vert_{r=1}=0$. Hence the first term in last
line of (21) is just a scalar heat kernel,
$$K_B(t)=K_N(S=-2,t)-\sum_{\lambda_0} \exp (-\lambda_0^2 t)
\eqno (22)$$
where $N$ stands for the Neumann boundary conditions.
The same procedure can be done for vectors $\phi_i$ [7]
satisfying Dirichlet boundary conditions.
$$K_\phi (t)=\sum_{l=1}^\infty 2((l+1)^2-1) \exp (-\lambda_l^2t)=$$
$$=\sum_{l=0}^\infty 2((l+1)^2-1) \exp (-\lambda_l^2t)=$$
$$=2\sum_{l=0}^\infty (l+1)^2 \exp (-\lambda_l^2t)
-[2\sum_{n=1}^\infty \sum_{\lambda_n} \exp (-\tilde \lambda_n^2
t) +\sum_{\tilde \lambda_0} \exp (-\tilde \lambda_0^2t)]+$$
$$+\sum_{\tilde \lambda_0} \exp (-\tilde \lambda_0^2t)
\eqno (23)$$
where we changed the summation index $n=l+1$ and introduced tilde
over $\lambda_n$. The first sum is just the heat kernel $K_D(t)$
for scalar fields satisfying Dirichlet boundary conditions.
The terms in square brackets can be identified with the
scalar heat kernel on two-dimensional unit disk. Making use
of the identity $(\partial_0 +\frac 1r )J_1=J_0$, we obtain
$\tilde \lambda_0 =\lambda_0$, with $\lambda_0$ from the
equations (21) and (22).
Hence,
$$K_B(t)+K_\phi (t) =K_N(S=-2,t)+2K_D(t)-K_D(d=1,t)
\eqno (24)$$
For the heat kernels in $d=3$ the dimensionality is not
shown manifestly.

The contribution of the field $A_i^T$ can be evaluated in
a similar way
$$K_A(t)=2K_N(S=-1,t)-K_N(S=0,d=1,t)+
\sum_{\kappa} \exp (-\kappa^2 t), \eqno (25)$$
where $\kappa$ is defined by the condition
$\partial_0J_0 (\kappa )=0$.

The contribution of the scalar field $\psi$ is the standard scalar
heat kernel up to the subtraction of the $l=0$ modes which do not
generate any 3-dimensional vector fields. For $l=0$ the condition
$J_1(\kappa )=0$ selects the same eigenvalues as in eq (25),
$\partial_0 J_0 =-J_1$. However, now we should not exclude
the constant mode $\kappa =0$, because it is already excluded
by the Dirichlet boundary condition.
$$K_\psi (t)=K_D(t)-\sum_\kappa \exp (-\kappa^2t)+1 \eqno (26)$$

Collecting together eqs. (24)-(26) one obtains the heat kernel
for two-form satisfying absolute boundary condition
$$K(A;t)=3K_D(t)+K_N(S=-2,t)+2K_N(S=-1,t)-$$
$$-K_D(d=1,t)-
K_N(S=0,d=1,t)+1 \eqno (27)$$
By repeating this procedure for relative boundary conditions we
obtain
$$K(R;t)=K(A;t) \eqno (28)$$
The relation (28) was obvious before any calculations because
relative boundary conditions are dual to the absolute ones.
For the two-forms in $d+1=4$ the duality transformation maps the
functional space on itself. One can consider (28) as a consistency
check.

One can evaluate the $\zeta (0)$ for two-form by using the expressions
for the heat kernel expansion in the case of pure boundary conditions.
We obtain $\zeta (0)=\frac 7{15}$. This value agrees with calculations
[6] based on analytic formulae [8,7] for mixed boundary conditions.

In a conclusion, let us formulate main results of this work. We suggested
a method of the reduction of the eigenvalue problem for the Laplace
operator acting on $p$-forms obeying mixed boundary conditions
to the eigenvalue problem for pure boundary conditions.
The case of rank two forms was studied in some detail. We expressed
the tensor harmonics
on a disk via transverse tensor, transverse vector and scalar
harmonics on $S^d$. A complete analysis was performed for $d=3$.
We express the heat kernel as a whole in terms of heat kernels
for scalar fields satisfying pure boundary conditions.
In a particular case of $\zeta (0)$ we find complete agreement
with previous calculations [6]. Our results are especially
efficient for higher coefficients of the Seeley-Gilkey expansion
and can be applied to other fields [1,3-6,9].

\vspace{3mm}

{\bf Acknowledgements}

I am indebted to Peter Gilkey for sending me preliminary results
of his computations of the coefficient $a_5$ on manifolds with
boundaries. This work was inspired by these computations and
was originally intended for cross-checking. I am also grateful to
Giampiero Esposito for fruitful discussions. I would like to
thank Professor Abdus Salam, IAEA and UNESCO for hospitality
at the International Centre for Theoretical Physics, Trieste.
This work was partially supported by the Russian Foundation
for Fundamental Studies, grant 93-02-14378.

\vspace{3mm}

{\bf REFERENCES}
\newline
[1] G.Esposito, Quantum gravity, quantum cosmology and Lorentzian
geometries (Lecture Notes in Physics m12, Berlin, Springer, 1992).
\newline
[2] P.B.Gilkey, Invariance theory, the heat equation and the
Atiyah-Singer theorem (Publish or Perish, Delaware, 1984).
\newline
[3] I.G.Moss and S.J.Poletti, Phys. Lett. B245 (1990) 355; Nucl.
Phys. B341 (1990) 155;

S.J.Poletti, Phys. Lett. B249 (1990) 355.
\newline
[4] H.C.Luckock and I.G.Moss, Class. Quantum Grav. 6 (1989) 1993;

H.C.Luckock, J. Math. Phys. 32 (1991) 1755.
\newline
[5] G.Esposito, Nuovo Cimento 109B (1994) 203;

G.Esposito, A.Yu.Kamenshchik, I.V.Mishakov and G.Pollifrone,
DSF preprint 94/4 (1994), to appear in Class. Quantum Grav.
\newline
[6] I.G.Moss and S.Poletti, Phys. Lett. B333 (1994) 326.
\newline
[7] D.V.Vassilevich, Vector fields on a disk with mixed boundary
conditions, St.Petersburg preprint SPbU-94-6, gr-qc/9404052, to
be published.
\newline
[8] T.Branson and P.B.Gilkey, Commun. Part. Diff. Eqs. 15 (1990) 245.
\newline
[9] P.D.D'Eath and G.Esposito, Phys. Rev. D43 (1991) 3234; D44 (1991)
1713;

A.Yu.Kamenshchik and I.V.Mishakov, Phys. Rev. D47 (1993) 1380;

A.O.Barvinsky, A.Yu.Kamenshchik and I.P.Karmazin, Ann. Phys. N.Y.
219 (1992) 201;

G.Esposito, A.Yu.Kamenshchik, I.V.Mishakov and G.Pollifrone,
DSF preprint 92/14, to appear in Phys. Rev. D.

\end{document}